\documentclass[12pt]{article}
\usepackage{epsfig}
\title{ \bf Leptonic Decay Constants of $D_{s}$ and $B_{s}$ Mesons at Finite
Temperature}

\author{El\c{s}en Veli Veliev *, G\"{u}l\c{s}ah Kaya **
\\Physics Department, Kocaeli University, Umuttepe Yerle\c{s}kesi \\
41380 Izmit, Turkey \\** e-mail: elsen@kocaeli.edu.tr
\\*** e-mail: gulsahbozkir@kocaeli.edu.tr}
\date{}
\begin{document}
\setlength{\baselineskip}{24pt}
\maketitle
\setlength{\baselineskip}{7mm}
\begin{abstract}
In the present work, $D_{s}$ and $B_{s}$ meson parameters are
investigated in the framework of thermal QCD sum rules. The
temperature dependences of the mass and the leptonic decay constants
are investigated by using Borel transform sum rules and Hilbert
moment sum rules. To increase sensitivity, the vacuum contributions
are subtracted from thermal expressions and the temperature
dependences of the leptonic decay constants and meson masses are
studied.
\end{abstract}

\setcounter{page}{1}
\section{Introduction}
In order to explain the heavy ion collision results, some
information about hadrons parameters at finite temperature and
density is required. Some of the characteristic parameters at finite
temperature and density are the masses and leptonic decay constants
of hadrons. The investigation of these parameters requires
non-perturbative approaches. One of these non-perturbative methods
is the QCD sum rules \cite{1}, formulated by Shifman, Vainshtein and
Zakharov.

The extension of the QCD sum rules method to finite temperatures has
been made by Bochkarev and Shaposhnikov \cite{2}. This extension is
based on two basic assumptions that the OPE and notion of
quark-hadron duality remain valid, but the vacuum condensates are
replaced by their thermal expectation values. The thermal QCD sum
rules method has been extensively used for studying thermal
properties of both light and heavy hadrons as a reliable and
well-establish method \cite{3}-\cite{8}.

The investigation of heavy meson decay constants at zero temperature
has been widely discussed in the literature \cite{9}. The knowledge
of these constants is needed in order to predict numerous heavy
flavor electroweak transitions and to determine Standard Model
parameters from the experimental data. Also leptonic decay constants
play essential role in the analysis of CKM matrix, CP violation and
the mixings $\overline{B_{d}}B_{d}$, $\overline{B_{s}}B_{s}$ . The
first determination of these constants were made twenty years ago
\cite{10}-\cite{12} and due to further theoretical and experimental
progress, this problem was reconsidered taking into account the
running quark masses and perturbative three-loop $\alpha_{s}^{2}$
corrections to the correlation function \cite{13}, \cite{14}. At
finite temperatures the nonperturbative nature of QCD vacuum induces
temperature dependences of the leptonic decay constants and masses.
Recently, first attempts have been made in order to calculate the
leptonic decay constants of heavy mesons at finite temperature in
the framework of thermal QCD sum rules \cite{15}.

In the present paper, we investigate the temperature behavior of the
masses and leptonic decay constants of $D_{s}$ and $B_{s}$ mesons
using QCD sum rules. Taking into account perturbative two-loop order
$\alpha_{s}$ corrections to the correlation function and
nonperturbative corrections up to the dimension six condensates
\cite{16} we investigated the temperature dependences of masses and
leptonic decay constants using Borel transform sum rules and Hilbert
moment sum rules. For increased sensitivity, we subtract the vacuum
contributions from thermal expressions and study the temperature
dependences of the leptonic decay constants and meson masses.

\section{Pseudoscalar thermal correlator at finite temperature}
We start with pseudoscalar two-point thermal correlator
\begin{equation}\label{eqn1}
\psi_{5}(q^{2})=i \int d^{4}x e^{iq\cdot x} \langle T(J(x)J^{+}(0))\rangle, \\
\end{equation}
where $J(x)=(m_{Q}+m_{s}):\bar{s}(x)i\gamma_{5}Q(x):$ is the
heavy-light quark current and has the quantum numbers of the $D_{s}$
and $B_{s}$ mesons, $m_{Q}$ and $m_{s}$ are heavy and strange quark
masses respectively. We shall not neglect $s$ quark mass throughout
this work. Thermal average of any operator \emph{O} is defined in
the following way
\begin{equation}\label{eqn2}
\langle O\rangle=Tr e^{-\beta H}O/Tr e^{-\beta H}, \\
\end{equation}
where $H$ is the QCD Hamiltonian,  $\beta=1/T$ stands for the
inverse of the temperature $T$  and traces are over any complete set
of states. Up to a subtraction polynomial, which depends on the
large $q^{2}$ behavior, $\psi_{5}(q^{2})$  satisfies the following
dispersion relation \cite{1},\cite{9}
\begin{equation}\label{eqn3}
\psi_{5}(Q^{2})=\int ds \frac{\rho (s)} {s+Q^{2}}+ subtractions , \\
\end{equation}
where $Q^{2}=-q^{2}$ is Euclidean momentum, $\rho
(s)=\frac{1}{\pi}Im \psi_{5}(s)$ is spectral density and in
perturbation theory at zero temperature in the leading order has the
following form \cite{16}:
\begin{equation}\label{eqn4}
\rho (s)=\frac{3(m_{s}+m_{Q})^{2}}{8\pi^{2}s}v(s)
q^{2}(s)\Big{[}1+\frac{4\alpha_{s}}{3\pi}f(x)\Big{]}, \\
\end{equation}
where $x={m_{Q}^{2}}/{s}$, $\alpha_{s}=\alpha_{s}(m_{Q}^{2})$  and
\begin{equation}\label{eqn5}
q(s)=s-(m_{Q}-m_{s})^{2},~~v(s)=(1-4m_{s}m_{Q}/q(s))^{1/2}, \\
\end{equation}
\begin{equation}\label{eqn6}
f(x)=\frac {9}{4}+2Li_{2}(x)+\ln {x}\ln {(1-x)}-\frac {3}{2}\ln
{\Big{(}\frac {1}{x}-1\Big{)}}-\ln {(1-x)}+x\ln {\Big{(}\frac
{1}{x}-1\Big{)}}-\frac{x}
{1-x}\ln {x}. \\
\end{equation}
The subtraction terms are removed by using the Borel transformation
or Hilbert moment methods. Therefore we will omit these terms. The
thermal propagator contains on-shell soft quarks which do not exist
in the confined phase. Therefore, in obtaining the OPE of the
thermal correlator (1), vacuum propagators must be used \cite{4}.
The non-perturbative contributions at zero temperature to the
correlator has the following form
\begin{eqnarray}\label{eqn7}
&&\psi_{5,np}(Q^{2})=-m_{Q} \lambda
\langle0|\bar{s}s|0\rangle\Big{[}1+\frac
{1}{2}\varepsilon(3-\lambda)-\lambda
\varepsilon^{2}(1-\lambda)-\frac
{1}{2}\varepsilon^{3}(1+\lambda-4\lambda^{2}+2 \lambda^{3})\Big{]}
\nonumber\\&&+\frac{1} {12\pi}
\lambda\langle0|\alpha_{s}G^{2}|0\rangle\Big{[}1+3\varepsilon\Big{(}1-\frac{8}
{3}\lambda+2\lambda^{2}-2\lambda(1-\lambda)\ln
{(\varepsilon\lambda)}\Big{)}\Big{]}\nonumber\\&&-\frac {M_{0}^{2}}
{2m_{Q}}
\langle0|\bar{s}s|0\rangle\lambda^{2}(1-\lambda)(1+\varepsilon(2-\lambda))-\frac{8\pi\rho}{27m_{Q}^{2}}\alpha_{s}\langle0|\bar{s}s|0\rangle^{2}\lambda^{2}(2-\lambda-\lambda^{2})
,\end{eqnarray}
where $\lambda=m_{Q}^{2}/(Q^{2}+m_{Q}^{2})$ and
$\varepsilon=m_{s}/m_{Q}$. Also, for the mixed condensate the
parameterization:
\begin{equation}\label{eqn8}
g\langle0|\bar{q}\sigma_{\mu\nu}\frac {\lambda_{a}} {2}
G_{a}^{\mu\nu}q|0\rangle=M_{0}^{2}\langle0|\bar{q}q|0\rangle
\\
\end{equation}
is used. It is assumed, that the expansion (7) also remains valid at
finite temperatures, but the vacuum condensates must be replaced by
their thermal expectation values \cite{2}. For the light quark
condensate at finite temperature we use the results of \cite{17},
\cite{18} obtained in chiral perturbation theory. Temperature
dependence of quark condensate in a good approximation can be
written as
\begin{equation}\label{eqn9}
\langle\bar{q}q\rangle=\langle0|\bar{q}q|0\rangle\Big{[}1-0.4\Big{(}\frac{T}
{T_{c}}\Big{)}^{4}-0.6\Big{(}\frac{T} {T_{c}}\Big{)}^{8}\Big{]}, \\
\end{equation}
where $T_{c}=160~MeV$ is the critical temperature. The low
temperature expansion of the gluon condensate is proportional to the
trace of the energy momentum tensor \cite{19} and can be
approximated by \cite{15}
\begin{equation}\label{eqn10}
\langle\alpha_{s}G^{2}\rangle=\langle0|\alpha_{s}G^{2}|0\rangle
\Big{[}1-\Big{(}\frac{T} {T_{c}}\Big{)}^{8}\Big{]}
. \\
\end{equation}
The value of the QCD scale $\Lambda$ is extracted from the value of
$\alpha_{s}(M_{Z})=0.1176$ \cite{20}. Equating OPE and hadron
representations of the correlation function and using quark-hadron
duality the sum rules is obtained as
\begin{equation}\label{eqn11}
\frac {f_{H}^{2}m_{H}^{4}} {Q^{2}+m_{H}^{2}}=
\int^{s_{0}}_{(m_{Q}+m_{s})^{2}} ds \frac {\rho(s)}
{s+Q^{2}}+\psi_{5,np}(Q^{2}),
\\
\end{equation}
where $f_{H}$ is the leptonic decay constant and is defined by the
matrix element of the axial-vector current between the corresponding
meson and the vacuum as:
\begin{equation}\label{eqn12}
\langle0|\bar{s}\gamma_{\mu}\gamma_{5}Q|H(q)\rangle=if_{H}q_{\mu},
\\
\end{equation}
where $Q=c,b$ and $H=D_{s},B_{s}$ in the same normalization as
$f_{\pi}=130.56~MeV$. In thermal field theories the parameters
$m_{H}$ and $f_{H}$ must be replaced by their temperature dependent
values. The continuum threshold $s_{0}$ also depends on temperature;
to a very good approximation it scales universally as the quark
condensate \cite{15}
\begin{equation}\label{eqn13}
s_{0}(T)=s_{0}\frac{\langle\bar{q}q\rangle}
{\langle0|\bar{q}q|0\rangle}\Big{[}1-\frac{(m_{Q}+m_{s})^{2}}{s_{0}}\Big{]}+(m_{Q}+m_{s})^{2},
\\
\end{equation}
where in the right hand side $s_{0}$ is hadronic treshold at zero
temperature: $s_{0}\equiv s_{0}(0)$. Analysis shows that thermal
non-perturbative correlator is basically driven by the quark
condensates.
\section{Numerical analysis of masses and leptonic decay constants}
In this section we present our results for the temperature
dependence of $D_{s}$ and $B_{s}$ meson masses and leptonic decay
constants. Performing Borel transformation with respect to
$Q_{0}^{2}$ on both sides of equation (11) and differentiating with
respect to $1/M^{2}$, we obtain:
\begin{equation}\label{eqn14}
m_{H}^{2}(T)=\frac{f_{H}^{2}m_{H}^{6}exp(-m_{H}^{2}/M^{2})+\overline{B}(T)}{f_{H}^{2}m_{H}^{4}exp(-m_{H}^{2}/M^{2})+\overline{A}(T)},\\
\end{equation}
\begin{equation}\label{eqn15}
f_{H}^{2}(T)=\frac{1}{m_{H}^{4}(T)}\Big{[}\overline{A}(T)+f_{H}^{2}m_{H}^{4}exp\Big{(}-\frac{m_{H}^{2}}{M^{2}}\Big{)}\Big{]}exp\Big{[}\frac{m_{H}^{2}(T)}{M^{2}}\Big{]},\\
\end{equation}
where the bar on the operators means subtractions of their vacuum
expectation values from thermal expectation values;  for example
$\overline{\psi_{5,np}}(M^{2},T)=\psi_{5,np}(M^{2},T)-\psi_{5,np}(M^{2},T=0)$.
Here
\begin{equation}\label{eqn16}
\overline{A}(T)=\int^{s_{0}(T)}_{s_{0}}
ds\rho(s)exp\Big{(}-\frac{s}{M^{2}}\Big{)}+\overline{\psi_{5,np}}(M^{2},T),\\
\end{equation}
\begin{eqnarray}\label{eqn17}
&&\overline{\psi_{5,np}}(M^{2},T)=-m_{Q}^{3}\overline{\langle0|\overline{s}s|0\rangle}e^{-\beta}\Big{[}1+\frac{3}{2}\varepsilon-\frac{1}{2}\varepsilon\beta-\beta\varepsilon^{2}\Big{(}1-\frac{1}{2}\beta\Big{)}-\frac{1}{2}\varepsilon^{3}\Big{(}1+\beta-2\beta^{2}+\frac{1}{3}\beta^{3}\Big{)}\Big{]}
\nonumber\\&&+\frac{1}{12\pi}\overline{\langle0|\alpha_{s}G^{2}|0\rangle}
m_{Q}^{2}e^{-\beta}\Big{[}1+3\varepsilon\Big{(}1-\frac{8}{3}\beta+\beta^{2}-2\beta(\ln(\beta\varepsilon)+\gamma-1)+\beta^{2}\Big{(}\ln(\beta\varepsilon)+\gamma-\frac{3}{2}\Big{)}\Big{)}\Big{]}
\nonumber\\&&-\frac{1}{2}M_{0}^{2}m_{Q}\beta\overline{\langle0|\overline{s}s|0\rangle}e^{-\beta}\Big{[}1-\frac{1}{2}\beta+2\varepsilon\Big{(}1-\frac{3}{4}\beta\Big{(}1-\frac{1}{9}\beta\Big{)}\Big{)}\Big{]}-\frac{4}{81}\pi\rho\alpha_{s}\overline{\langle0|\bar{s}s|0\rangle^{2}}\beta
e^{-\beta}\nonumber\\&&\times(12-3\beta-\beta^{2}),\end{eqnarray}
where $\gamma$ is the Euler constant, $\beta=m_{Q}^{2}/M^{2}$ and
$\overline{B}(T)=-m_{Q}^{2}\frac{d\overline{A}(T)}{d\beta}$. To
investigate the meson parameters at finite temperature we also use
Hilbert moments methods, which eliminate the subtraction terms.
Calculating Hilbert moments at $Q^{2}=-q^{2}=0$ and using first two
moments we obtain
\begin{equation}\label{eqn18}
m_{H}^{2}(T)=\frac{F(T)-\int^{s_{0}}_{s_{0}(T)} ds
\rho(s)s^{-3}+f_{H}^{2}/m_{H}^{2}}{G(T)-\int^{s_{0}}_{s_{0}(T)} ds
\rho(s)s^{-4}+f_{H}^{2}/m_{H}^{4}},\\
\end{equation}
\begin{equation}\label{eqn19}
f_{H}^{2}(T)=m_{H}^{2}(T)\big {[}F(T)-\int^{s_{0}}_{s_{0}(T)} ds
\rho(s)s^{-3}+f_{H}^{2}/m_{H}^{2}\big {]},\\
\end{equation}
where $F(T)$ and $G(T)$ functions are expressed by thermal
expectation values of condensates
\begin{eqnarray}\label{eqn20}
&&F(T)=-\frac{1}{m_{Q}^{3}}\overline{\langle0|\bar{s}s|0\rangle}(1+3\varepsilon^{2})+\frac{1}{12\pi
m_{Q}^{4}}\overline{\langle0|\alpha_{s}G^{2}|0\rangle}[1+\varepsilon(21+18\ln\varepsilon)]\nonumber\\&&+\frac{1}{2m_{Q}^{5}}M_{0}^{2}\overline{\langle0|\overline{s}s|0\rangle}(3+2\varepsilon)+\frac{80}{27m_{Q}^{6}}\pi\rho\alpha_{s}\overline{\langle0|\bar{s}s|0\rangle^{2}},\end{eqnarray}
\begin{eqnarray}\label{eqn21}
&&G(T)=-\frac{1}{m_{Q}^{5}}\overline{\langle0|\overline{s}s|0\rangle}\Big{(}1-\frac{1}{2}\varepsilon+6\varepsilon^{2}-\frac{5}{2}\varepsilon^{3}\Big{)}+\frac{1}{12\pi
m_{Q}^{6}}\overline{\langle0|\alpha_{s}G^{2}|0\rangle}[1+\varepsilon(52+36\ln\varepsilon)]\nonumber\\&&+\frac{1}{m_{Q}^{7}}M_{0}^{2}\overline{\langle0|\overline{s}s|0\rangle}(3+\varepsilon)+\frac{176}{27m_{Q}^{8}}\pi\rho\alpha_{s}\overline{\langle0|\bar{s}s|0\rangle^{2}}.\end{eqnarray}
\begin{table}[h]
\begin{center}
\caption{QCD input parameters used in the analysis.}
\begin{tabular}{ll}
\\
\\
\hline
  \\
Parameters& References\\
\\
\hline
\\
$m_{D_{s}}=1968$ MeV&\cite{20}\\
$m_{B_{s}}=5366$ MeV&\cite{20}\\
$m_{s}=120$ MeV &\cite{20}\\
$m_{c}=1.47$ GeV&\cite{13,20}\\
$m_{b}=4.4$ GeV&\cite{13,20}\\
$f_{D_{s}}=235$ MeV&\cite{13,20}\\
$f_{B_{s}}=240$ MeV&\cite{13,20}\\
$\rho=4$&\cite{15,16}\\
$\langle0|\overline{q}q|0\rangle=-0.014~$GeV$^3$&\cite{1}\\
$\langle0|\frac {1}{\pi}\alpha_{s}G^{2}|0\rangle=0.012~$GeV$^4$&\cite{1}\\
$\alpha_s\langle0|\overline{q}q|0\rangle^{2}=5.8\times10^{-4}~$GeV$^6$&\cite{13}\\
$M_{0}^{2}=0.8~$GeV$^2$&\cite{13}\\
$\langle0|\overline{s}s|0\rangle=0.8\langle0|\overline{q}q|0\rangle$&\cite{13}\\
\\
\hline
\end{tabular}
\end{center}
\end{table}
For the numerical evolution of the above sum rule, the values of the
QCD parameters used are shown in Table 1. The criterion we adopt
here is to fix $s_{0}$ in such a way as to reproduce the zero
temperature values of meson masses and leptonic decay constants. For
$D_{s}$ meson $s_{0}$ is $6~GeV^2$ and $8~GeV^2$, for $B_{s}$ meson
$s_{0}$ is $34~GeV^2$ and $35~GeV^2$ in Borel and Hilbert moment sum
rules methods, respectively.
The temperature dependences of the $D_{s}$ and $B_{s}$ meson masses
and leptonic decay constants obtained using the Borel and Hilbert
moment methods are shown in Fig. 1 and Fig. 2, respectively. The
results for leptonic decay constants are shown in Fig. 3 and Fig. 4.
As seen in figures, $f_{D_{s}}$ and $f_{B_{s}}$ decrease with
increasing temperature and vanish approximately at critical
temperature $T_{c}=160~MeV$. This may be interpreted as a signal for
deconfinement and agrees with light and heavy-light mesons
investigations \cite{15}, \cite{21}. Numerical analysis shows that
the temperature dependence of $f_{D_{s}}$ is independent of $M^{2}$,
when $M^{2}$ changes between $3~GeV^{2}$ and $4~GeV^{2}$ and
$f_{B_{s}}$ is independent of the Borel parameter, when $M^{2}$
changes between $16~GeV^{2}$ and $24~GeV^{2}$.
Obtained results can be used for the interpretation of heavy ion
collision experiments. It is also essential to compare these results
with other model calculations. We believe these studies to be of
great importance for understanding phenomenological and theoretical
aspects of thermal QCD.

\section{Acknowledgement}
The authors much pleasure to thank T. M. Aliev and A. \"{O}zpineci
for useful discussions. This work is supported by the Scientific and
Technological Research Council of Turkey (TUBITAK), research project
no.105T131.
\begin{figure}[h]
\centerline{\epsfig{figure=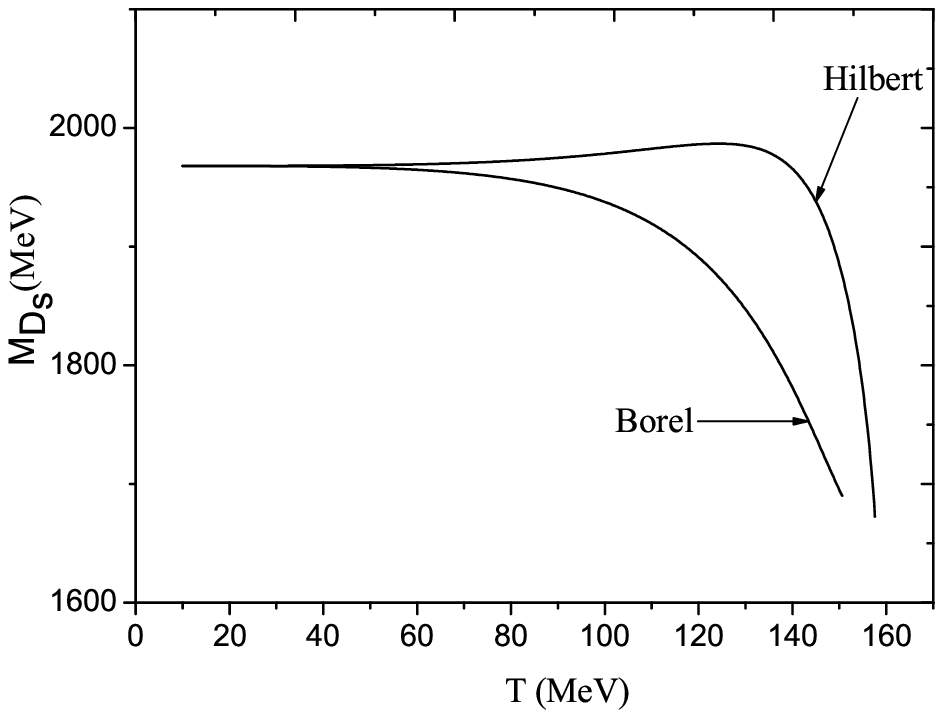,height=80mm}}\caption{Temperature
dependence of $D_{s}$ meson mass in Hilbert and Borel sum rules
methods. Here Borel parameter is $M^{2}=3~GeV^2$, hadronic threshold
$s_{0}=6~GeV^2$ for Borel and
~~~~~~~~~~~~~~~~~~~~~~~$s_{0}=8~GeV^2$ for Hilbert moment sum rules
methods.} \label{BRmh0}
\end{figure}
\begin{figure}[h]
\centerline{\epsfig{figure=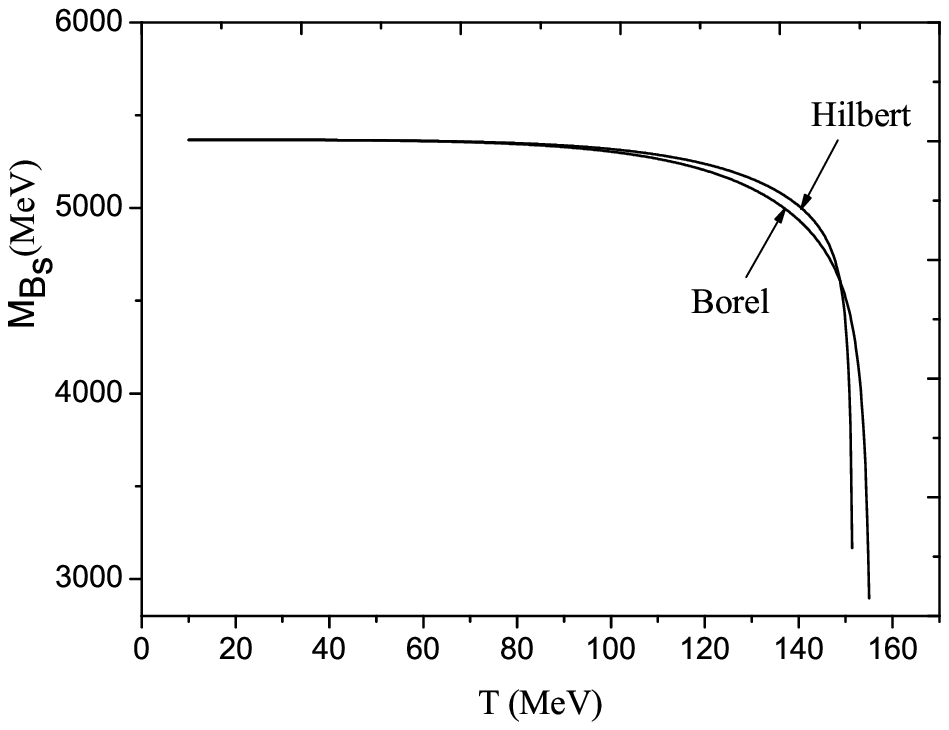,height=80mm}}\caption{
Temperature dependence of $B_{s}$ meson mass in Hilbert and Borel
sum rules methods. Here Borel parameter is $M^{2}=20~GeV^2$,
hadronic threshold $s_{0}=34~GeV^2$ for Borel and $s_{0}=35~GeV^2$
for Hilbert moment sum rules methods.} \label{BRmh0}
\end{figure}
\begin{figure}[h]
\centerline{\epsfig{figure=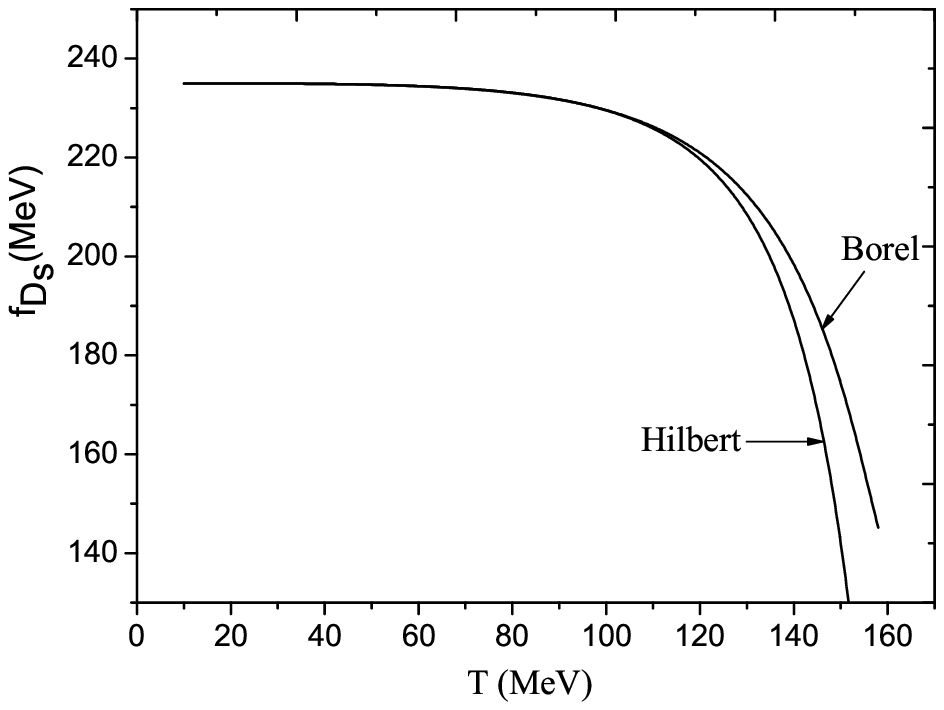,height=80mm}}\caption{
Temperature dependence of $f_{D_{s}}$ in Hilbert and Borel sum rules
methods. Here Borel parameter is $M^{2}=3~GeV^2$, hadronic threshold
$s_{0}=6~GeV^2$ for Borel and $s_{0}=8~GeV^2$ for Hilbert moment sum
rules methods.} \label{BRmh0}
\end{figure}
\begin{figure}[h]
\centerline{\epsfig{figure=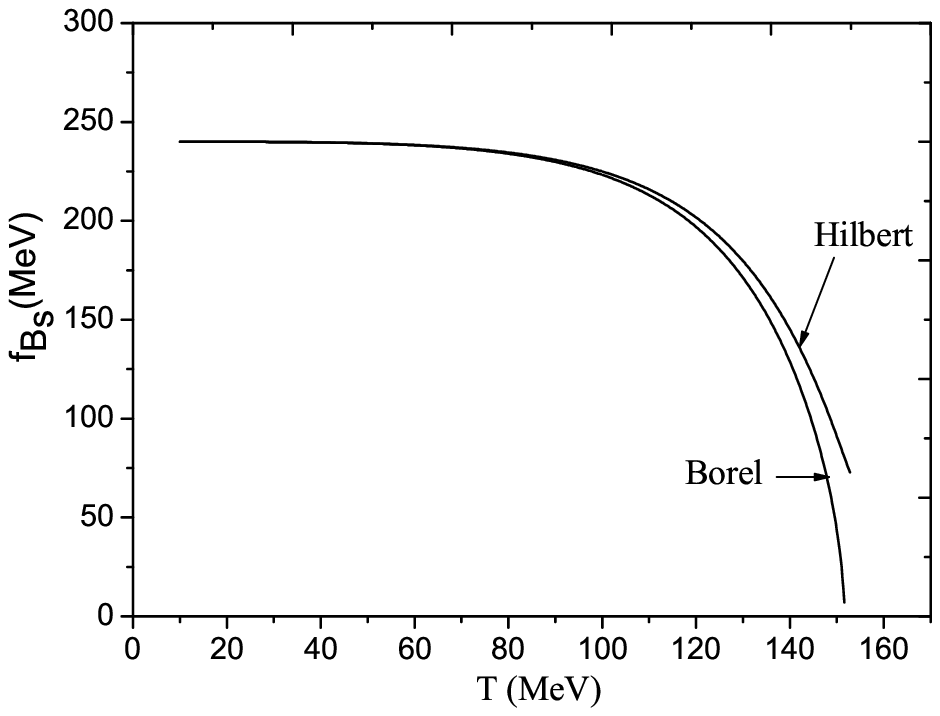,height=80mm}}\caption{
Temperature dependence of $f_{B_{s}}$ in Hilbert and Borel sum rules
methods. Here Borel parameter is $M^2=20~GeV^2$, hadronic threshold
$s_{0}=34~GeV^2$ for Borel and $s_{0}=35~GeV^2$ for Hilbert moment
sum rules methods.} \label{BRmh0}
\end{figure}
\end{document}